# DSPatch: Dual Spatial Pattern Prefetcher

Rahul Bera[1]    Anant V. Nori[1]    Onur Mutlu[2]    Sreenivas Subramoney[1]
[1]Processor Architecture Research Lab, Intel Labs    [2]ETH Zürich

## ABSTRACT

High main memory latency continues to limit performance of modern high-performance out-of-order cores. While DRAM latency has remained nearly the same over many generations, DRAM bandwidth has grown significantly due to higher frequencies, newer architectures (DDR4, LPDDR4, GDDR5) and 3D-stacked memory packaging (HBM). Current state-of-the-art prefetchers do not do well in extracting higher performance when higher DRAM bandwidth is available. Prefetchers need the ability to dynamically adapt to available bandwidth, boosting prefetch count and prefetch coverage when headroom exists and throttling down to achieve high accuracy when the bandwidth utilization is close to peak.

To this end, we present the Dual Spatial Pattern Prefetcher (DSPatch) that can be used as a standalone prefetcher or as a lightweight adjunct spatial prefetcher to the state-of-the-art delta-based Signature Pattern Prefetcher (SPP). DSPatch builds on a novel and intuitive use of modulated spatial bit-patterns. The key idea is to: (1) represent program accesses on a physical page as a bit-pattern anchored to the first "trigger" access, (2) learn two spatial access bit-patterns: one biased towards coverage and another biased towards accuracy, and (3) select one bit-pattern at run-time based on the DRAM bandwidth utilization to generate prefetches. Across a diverse set of workloads, using only 3.6KB of storage, DSPatch improves performance over an aggressive baseline with a PC-based stride prefetcher at the L1 cache and the SPP prefetcher at the L2 cache by 6% (9% in memory-intensive workloads and up to 26%). Moreover, the performance of DSPatch+SPP scales with increasing DRAM bandwidth, growing from 6% over SPP to 10% when DRAM bandwidth is doubled.

## 1 INTRODUCTION

High main memory latency continues to limit the performance of modern high-performance out-of-order (OOO) cores. Prefetching is a well-studied approach to mitigate the performance impact of the high memory latency [25, 29, 38, 43, 45, 49, 51, 54, 62, 65, 69, 72–74, 77]. A primary metric to improve performance using prefetchers is *coverage*, which is the fraction of program loads to memory that are removed by the prefetcher. At odds with coverage, but still very important, is prefetcher accuracy, which is the fraction of issued prefetches that are actually needed by the program loads. Inaccurate prefetches can pollute the small on-die caches and can cause excessive pressure on memory bandwidth, which in turn can increase the latency of responses from memory.

While DRAM latency has remained nearly the same over decades [27, 59], DRAM bandwidth has grown significantly [40, 58]. Higher DRAM core frequencies and new DRAM architectures (e.g., DDR4 [10], LPDDR4 [12], GDDR5 [11]) boost memory bandwidth at the same memory interface width. Newer 3D-stacked memory packages [7, 8, 24, 58] enable higher bandwidth by increasing the memory interface width. When higher DRAM bandwidth headroom is available, the negative impact due to inaccurate prefetches is lower. While the DRAM bandwidth is shared among multiple cores in a system, several prior studies across mobile, client and server systems [40, 47, 52, 61, 68, 75] have observed that the available memory bandwidth is not heavily utilized. Latency is often a bigger bottleneck than bandwidth, since either (1) there are very few active threads running in the system, (2) not all threads are memory sensitive, or (3) there is not enough memory parallelism present in the program to fully utilize the memory bandwidth [40]. Yet, as we demonstrate in Figure 1, current state-of-the-art prefetchers (Signature Pattern Prefetcher (SPP) [54], Best Offset Prefetcher (BOP) [62] and Spatial Memory Streaming (SMS) [73]) do not scale well in performance with increasing peak DRAM bandwidth. Prefetching techniques need to evolve to make the best use of this critical DRAM bandwidth resource when power budget is available. Specifically, prefetchers need to possess the ability to dynamically adapt to available DRAM bandwidth, boosting predictions and coverage when headroom exists and throttling down to achieve high accuracy when the bandwidth utilization is close to peak.

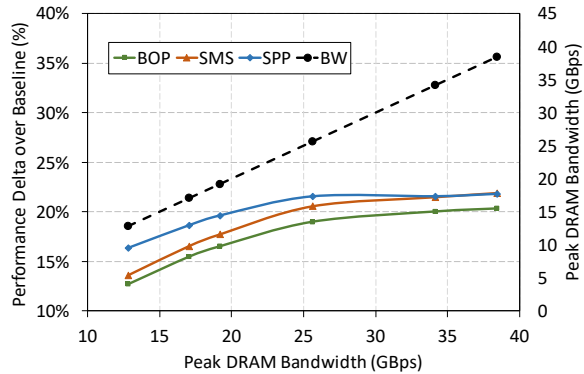

**Figure 1: Prefetcher performance scaling with DRAM bandwidth (points correspond to single and dual channels of DDR4-1600, 2133 and 2400)**

Prefetching is a speculation mechanism to predict future addresses to be accessed by the program. Address access patterns can be represented in various forms, including full addresses, offsets in a spatial region (typically a 4 KB page), or address deltas between accesses. Choosing an address access representation that has the best chance of exposing repeating patterns can help to boost prefetch coverage and performance.

Patterns in address accesses that are readily apparent when taking a global or accumulative view of accesses may not be visible when taking a restricted view of deltas between recent consecutive accesses. Figure 2 illustrates an example of multiple streams of accesses within a single spatial region and their representation in various formats. The first access to the region is called the "trigger" access. Access streams B through E have the same trigger offset in the spatial region and touch all the same offsets but in different temporal order. Such variations are typically an artifact of reordering due to out-of-order scheduling in the core and the cache/memory sub-systems. The longer the access sequence, the higher the probability of variations [43]. These access streams all have different representations when successive address deltas are used to represent the access patterns. Yet, we realize that they can actually be represented by a *single* spatial bit-pattern. For example, access streams B and C with trigger offset 1 have two different delta representations (+4,-1,+7,+1 and +4,+6,-7,+8) but the same (i.e., single) bit-pattern representation BP2 (0100 1100 0001 1000). Crucially, we observe that *when bit-patterns are anchored to the "trigger" offset (rotated left in this case), all access streams in the example can be represented by a single anchored bit-pattern.* Such an anchored bit-pattern essentially represents two views of the delta stream: the deltas between consecutive accesses (we call this the *local* view of deltas) and the deltas relative to the trigger access (we call this the global view of deltas).

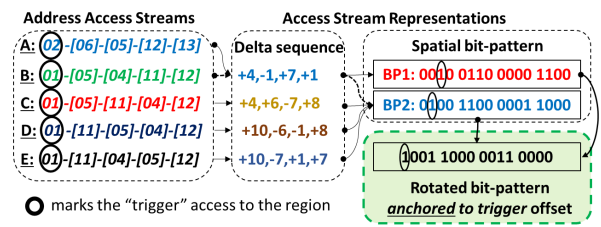

**Figure 2: Multiple different address access streams in a single memory region can be represented by a single spatial bit-pattern, anchored to their respective trigger accesses.**

While the use of anchored spatial bit-patterns can boost coverage, it does not provide the ability to adapt and scale prefetch coverage based on the available resources and DRAM bandwidth in the system. Multiple access streams in a spatial region can have anchored bit-patterns that are very similar (i.e., have some common set bits) but not exactly the same. We propose a novel and intuitive approach to simultaneously optimize coverage and accuracy by learning *two* bit-patterns: one biased towards coverage and another towards accuracy. A bitwise OR operation on the recently-observed anchored bit-patterns in a given memory region adds bits into a resultant bit-pattern (called the *coverage-biased bit-pattern*) and thus modulates the resultant bit-pattern for higher coverage. Similarly, a bitwise AND operation on the recently-observed anchored bit-patterns in a given memory region subtracts bits away from a resultant bit-pattern (called the *accuracy-biased bit-pattern*) and thus modulates the second resultant bit-pattern for higher accuracy. Figure 3 shows an example of how multiple different address patterns to the same memory region that map to three different anchored bit-patterns can be modulated into a coverage-biased bit-pattern (shown in green) and an accuracy-biased bit-pattern (shown in red). As we will show, dynamic modulation of these bit-patterns enables *simultaneous optimization for both coverage and accuracy*, even though these metrics are at odds with each other. The available memory bandwidth headroom, coupled with a quantified measure of accuracy and coverage can be used to select between the two two modulated bit-patterns dynamically at run-time.

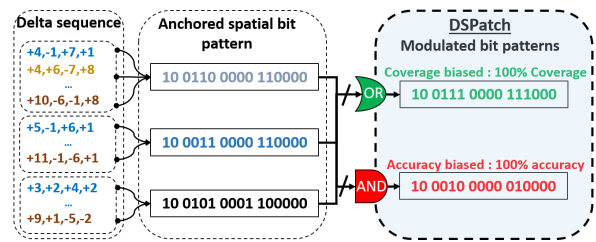

**Figure 3: Illustration of two modulated bit-patterns that DSPatch takes advantage of**

We present the Dual Spatial Pattern Prefetcher (DSPatch), a lightweight spatial prefetcher that can be used as a standalone prefetcher or as a light-weight adjunct spatial prefetcher to the state-of-the-art delta-based Signature Pattern Prefetcher (SPP). DSPatch builds on an intuitive use of simple logical OR and AND operations to learn two modulated spatial bit-pattern representations of accesses to a given memory region (i.e., a physical page). One bit-pattern is biased towards coverage and the other bit-pattern is biased towards accuracy. DSPatch employs a simple but effective method to track coverage and accuracy characteristics of each modulated bit-pattern as well as the overall DRAM bandwidth utilization. Based on this tracking information, DSPatch dynamically selects one bit-pattern to generate prefetches. If the DRAM bandwidth utilization is high, DSPatch selects the accuracy-biased bit-pattern for prefetching. If the DRAM bandwidth utilization is low, DSPatch selects the coverage-biased bit-pattern if the bit-pattern has good enough accuracy, or the accuracy-biased bit-pattern otherwise.

We make the following key contributions in this work:

- We observe that even though peak DRAM bandwidth is growing with newer DRAM architectures and packages, state-of-the-art prefetcher performance does not scale well with the growing DRAM bandwidth.
- We show that a spatial bit-pattern representation anchored around a trigger access to a region can effectively capture all deltas (local and global) from the trigger. This transformation exposes similar patterns that are otherwise obfuscated to look different due to memory access reordering in the processor and the memory subsystem.
- We introduce a new prefetching algorithm that learns two modulated bit-patterns to prefetch in a given memory region by using simple logical OR and AND operations. One bit-pattern is biased towards coverage and the other is biased towards accuracy.
- We propose a simple method to track DRAM bandwidth utilization and to measure the coverage and accuracy of

modulated bit-patterns. We show that this method enables effective dynamic selection of a single bit-pattern to generate prefetch candidates at run-time.

Across a diverse set of 75 workloads, with only 3.6KB of storage, DSPatch improves performance over an aggressive baseline that employs a PC-based stride prefetcher at the L1 cache and the SPP at the L2 cache by 6% (9% in memory-intensive workloads and up to 26%). As a standalone prefetcher, DSPatch has slightly (1%) higher performance than the state-of-the-art SPP with only $2/3^{rd}$ of the storage requirements of SPP. We find that, the use of SPP and DSPatch together combines the benefits of both the state-of-the-art fine-grained delta-based prefetching and the state-of-the-art bit-pattern-based prefetching. We show that, by simultaneously optimizing for both coverage and accuracy, every 2% increase in coverage with DSPatch comes at only a 1% increase in mispredictions. Finally, the performance of DSPatch+SPP scales well with increasing memory bandwidth, growing from 6% over SPP to 10% when DRAM bandwidth is doubled.

## 2 BACKGROUND AND MOTIVATION

Prefetching is a speculation technique that predicts the addresses of high-latency accesses in the program and brings the associated data into low-latency on-die caches for use by the program. High-latency accesses, typically to DRAM main memory, often stall the retirement of instructions in a core [65, 66]. They also reduce the look-ahead for instruction-level-parallelism (ILP) extraction, since filling up of the re-order buffer (ROB) due to memory access related stalls prevents allocation of younger independent instructions into processor structures [65, 66].

Multiple mechanisms to represent address access patterns in a program have been studied over the years. Address access patterns can be represented via various means, including (1) full addresses, (2) offsets in a spatial region (typically a 4 KB page) or (3) address deltas between consecutive accesses. In order to identify such patterns, prefetchers typically examine a subset of memory accesses filtered by certain program context. For example, a PC-based stride prefetcher tries to learn a constant stride between consecutive cacheline addresses referenced by a program counter (PC) value. Here, the PC acts as a program context that filters out accesses, in order to easily discover the access pattern. We call this program context that is used for filtering accesses as *signature*. A signature can be constructed using memory access information of a program like physical addresses, offsets or deltas between consecutive accesses, and can potentially be augmented with program control flow information like the PC. Prefetchers typically learn the repeating program address access pattern by correlating it with a signature and predict that the learnt pattern will be needed again when the signature is seen again. The design choice a prefetcher makes on what to use as a signature and as the address access pattern representation determines its effectiveness at optimizing the four main metrics of prefetching:

- Coverage: The fraction of high-latency memory accesses of the program that are saved by the prefetcher (the higher the better)
- Timeliness: The fraction of the latency of the high-latency accesses hidden by the prefetcher (the higher the better)

- Accuracy: The fraction of prefetched addresses that are later needed by the program (the higher the better)
- Storage: The hardware storage requirements of the prefetcher (the smaller the better)

In the rest of this section, we comprehensively examine three state-of-the-art prefetchers (SPP [54], BOP [62] and SMS [73]) with respect to their choices of signature and address access pattern representations. By analyzing a wide range of workloads, we compare the performance and identify the merits of the three types of prefetching. Finally, for each of the prefetchers, wherever possible, we evaluate potential dynamic bandwidth-aware tuning opportunities (similar to [35, 74]) for higher coverage to arrive at the prefetcher's best possible scalability in the presence of memory bandwidth headroom.

### 2.1 Signature Pattern Prefetcher (SPP)

SPP [54] is the state-of-the-art delta-based prefetcher. It uses a signature comprised solely of up to four recent consecutive address deltas (which we call "local" deltas) observed in a 4KB page. Each signature tracks at most four possible next deltas as prefetch candidates, along with a confidence value associated with each candidate. This allows SPP to track complex but repeating address delta patterns at low cost. SPP uses a recursive look-ahead mechanism to boost prefetch distance and timeliness. SPP appends each prefetch candidate delta recursively to the candidate delta's signature to generate further prefetch candidates. The confidence of each new prefetch candidate is a cascaded product of confidences leading to the candidate's level and the candidate's stored confidence value. A prefetch delta candidate whose cascaded confidence value is above a threshold value triggers a prefetch.

A big advantage of a delta-based prefetcher is that every access can participate in generating prefetches. This effectively allows multiple "bites" at the "apple" (coverage) to boost performance. With low storage requirements and through the use of cascaded confidence values, SPP has fine-grained control over coverage, timeliness and accuracy. As seen in Figure 4, SPP outperforms both BOP and SMS in six out of our nine workload categories, as well as on average.

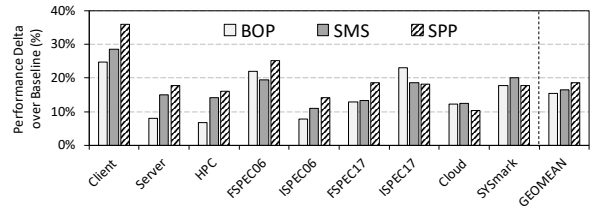

**Figure 4: Performance of BOP, SMS and SPP L2 prefetchers over a baseline with an L1 PC-stride prefetcher and a single channel of DDR4-2133**

However, there are scenarios where SPP loses out on coverage or timeliness. In pages with sparse and highly irregular access patterns, SPP cannot track all possible deltas, losing out on coverage. The deltas it tracks have low confidence values, limiting the recursive prefetch distance and hence timeliness. Due to these shortcomings,

SPP performs worse than either or both of the other two prefetchers in ISPEC17, Cloud and SYSMark workload categories.

***Bandwidth-aware tuning opportunity.*** SPP uses a static confidence threshold value of 25% to allow the prefetching of a candidate delta. With a simple dynamic scheme that monitors the available DRAM bandwidth headroom, we could modulate this threshold to lower values when DRAM bandwidth utilization is low. In Section 2.5 we evaluate an enhanced version of SPP (eSPP) that has the ability to lower its confidence threshold value to 12.5% if more than half of the DRAM bandwidth is not utilized. We observe that even eSPP shows poor performance scaling with higher memory bandwidth.

## 2.2 Best Offset Prefetcher (BOP)

BOP [62] is a delta-based prefetcher that aims to determine the set of optimal "global" deltas between accesses within a memory region (e.g., 4KB). For example, if a program experiences a repeating series of successive local deltas (1,2,1,2,1,2...), BOP identifies a single global delta of 3 (or its multiples) as an effective representation of address access patterns in the program. Further, BOP tracks and utilizes the most appropriate global delta to achieve timeliness (a multiple of 3 in this example).

Unlike SPP, which constructs a local view of accesses, BOP constructs a global view of accesses that helps BOP in two ways. First, the global view exposes more patterns in a memory region than a restricted local view of consecutive accesses. This especially helps BOP to predict future accesses in workloads with irregular access patterns with only few accesses per page. Second, the global view is robust against program access reordering, which can further disrupt the pattern learning based on a restricted local view of accesses. As seen in Figure 4, BOP, at a prefetch degree of two, has the highest performance among all prefetchers in the ISPEC17 workload category. However, BOP learns only a limited set of global deltas for *all* access streams in the program in a statically-defined epoch (defined by the number of accesses). This severely limits BOP's coverage and timeliness in HPC and Server workload categories.

***Bandwidth-aware tuning opportunity.*** The original BOP proposal tracks only a limited set of possible global deltas (in a 4KB page, 126 possible deltas from -63 to +63 exist) and statically picks a *single* best global delta per epoch (for a prefetch degree of one). In Section 2.5, we evaluate an enhanced bandwidth-aware version of BOP (called eBOP) that can adapt to DRAM bandwidth headroom. eBOP has a default prefetch degree of one, but can dynamically increase its degree to two and four if the bandwidth headroom is more than 25% and 50%, respectively. We find that neither BOP's nor eBOP's performance improvement scales well with additional memory bandwidth. This is because of two reasons. First, BOP's predictions suffer from poor accuracy since BOP does not use any program context information as a signature to generate prefetches. Second, any limit on prefetch degree hurts BOP's coverage.

## 2.3 Spatial Memory Streaming (SMS)

SMS [73] tracks address accesses within a spatial region (e.g., 2KB or 4KB) as a spatial bit-pattern. It maps each region's address access pattern to a signature comprising the *trigger* access PC and *trigger* offset in the region. The trigger access is defined as the first access to the region that adds the region to the structure that tracks recently-accessed regions. Doing so, SMS effectively exploits spatial correlations between an access from a PC and other accesses in the region. A bit-pattern representation inherently captures a global view of accesses in the region. Used in conjunction with a trigger PC based signature, SMS performs better than SPP in ISPEC17, Cloud and SYSmark workload categories (Figure 4).

However, SMS makes a number of static decisions that negatively affect its overall performance. Based on a study of available workloads [73], it statically decides to track 2KB regions (rather than 4KB) for accuracy reasons, thereby limiting prefetch distance and timeliness opportunities compared to SPP and BOP. With no explicit mechanism to track the accuracy of the stored bit-patterns, SMS relies on a high degree of access filtering through the use of sophisticated signature (PC+Offset). Therefore, to increase overall coverage, SMS relies on tracking a large number of signatures, increasing its storage requirements to tens of KB. Figure 5 shows that reducing the number of entries of pattern history table (i.e., the table that stores the correlation between signature and bit-pattern) in SMS from a baseline 16K entries (16-way associative) at 88KB storage down to 256 entries at 3.5KB storage approximately halves SMS's average performance improvement across all of our evaluated workloads. Furthermore, the current SMS design provides no clear opportunities to dynamically tune performance based on increase in DRAM bandwidth.

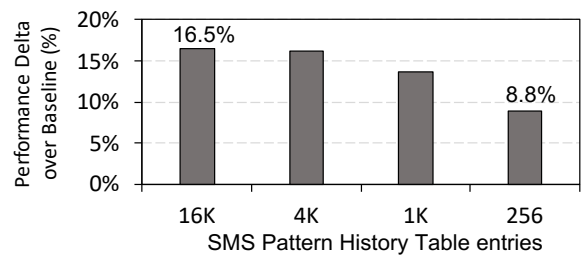

Figure 5: Performance impact of reducing SMS storage size from 16K entries at 88KB to 256 entries at 3.5KB, averaged across all workloads

## 2.4 Cache Pollution

Inaccurate prefetches can cause pollution in on-die caches by evicting useful cache blocks. The impact of pollution can be mitigated via the use of dead-block prediction [39, 48, 53] and prefetch-aware replacement and insertion policies [33–36, 46, 74, 79]. In fact, multiple generations of last-level cache replacement policies have specifically targeted identifying and replacing dead blocks [39, 53, 78]. We do not observe significant pollution impact of inaccurate prefetches. We find that the pressure on the memory bandwidth resource is the primary negative impact of inaccurate prefetches in the state-of-the-art prefetchers we examine.

## 2.5 Performance Scaling of Prefetchers with Memory Bandwidth Scaling

Figure 6 shows how the performance improvement of each prefetcher scales with increased memory bandwidth. We draw three major conclusions from the figure. First, the performance improvement of none of the three prefetchers scales well with increased memory bandwidth. In other words, the benefit of each prefetcher saturates as memory bandwidth increases. Second, the rate of the increase in performance improvement with increase in memory bandwidth is higher in SMS. In fact, the performance improvement of SMS matches that of SPP at higher memory bandwidth points. This shows the benefit of spatial bit-pattern prefetching in the presence of higher memory bandwidth. Third, eBOP enjoys the best performance scaling with memory bandwidth due to its dynamic modulation of the prefetch degree. However, the lack of program context information and the limited prefetching degree constrains eBOP coverage and leaves significant performance on the table.

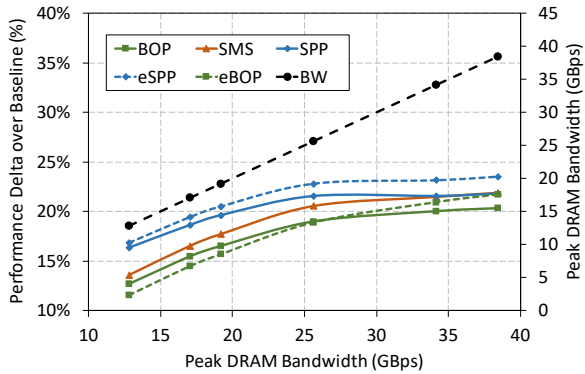

Figure 6: None of the five state-of-the-art prefetchers we examine, including eSPP and eBOP, scale well in performance with higher DRAM bandwidth.

## 2.6 Takeaways and Our Goal

In summary, our analysis of state-of-the-art prefetchers lead to three major takeaways:

- None of the state-of-the-art prefetchers we examine scale well in performance when higher DRAM bandwidth is available. SMS inherently lacks the ability to use available memory bandwidth in its algorithm to fine tune prefetch aggressiveness. SPP and BOP can be made bandwidth aware, yet they scale poorly in performance (as we see for eSPP and eBOP Figure 6).
- A spatial bit-pattern representation anchored around a trigger access to a memory region effectively captures all deltas in the region: local (deltas between consecutive accesses) and global (deltas with respect to the trigger access). This representation exposes patterns that are otherwise obfuscated by reordering in the processor and the memory subsystem.
- Using simple bit operations like OR and AND on the recently seen access bit-patterns in a memory region, we can learn two modulated bit-patterns, one biased towards coverage and the other biased towards accuracy. We can dynamically select the appropriate bit-pattern to generate prefetches to increase prefetch coverage when memory bandwidth utilization is low, or to increase prefetch accuracy when memory bandwidth utilization is high.

***Our goal*** is to design a spatial bit-pattern prefetcher that integrates the memory bandwidth inherently into its learning algorithm to dynamically adjust its notion of aggressiveness so that it can scale its performance improvement with higher memory bandwidth. To this end, we present the Dual Spatial Pattern Prefetcher (DSPatch), which makes use of the dual modulated spatial bit-patterns we introduced earlier to simultaneously optimize prefetch coverage and accuracy based on the memory bandwidth utilization.

## 3 DUAL SPATIAL PATTERN PREFETCHER

In this section, we describe the Dual Spatial Pattern Prefetcher (DSPatch), a spatial bit-pattern prefetcher that learns two bit-patterns per memory region (i.e., a physical page) and associates them with a program counter (PC) based signature:

- One bit-pattern (called *CovP*) is biased towards higher coverage. It is calculated as a simple OR of the recently observed spatial program access bit-patterns to the physical page. The OR operation adds bits to the learnt bit-pattern and grows the bit-pattern for higher coverage, up to a certain threshold.
- The other bit-pattern (called *AccP*) is biased towards higher accuracy. It is calculated as a simple AND of the coverage-biased bit-pattern (*CovP*) and the currently observed program access bit-pattern to the physical page. The AND operation reduces the set bits to maximize accuracy but since *AccP* is derived from *CovP*, coverage is kept in check.

DSPatch mainly comprises of two hardware structures: *Page Buffer* (*PB*) and *Signature Pattern Table* (*SPT*). The purpose of *PB* is to record the observed spatial bit-patterns as the program accesses a physical page. The purpose of *SPT* is to store the two modulated spatial bit-patterns (*CovP* and *AccP*), derived from previously-observed bit-patterns, by associating them with the trigger PC into the page. Thus, DSPatch observes program accesses per physical page, but learns overall program access patterns in a page-agnostic way by associating the spatial bit-patterns with the triggering PC signature. *SPT* is looked up with the triggering PC when a new physical page is accessed, to retrieve the two modulated bit-patterns: *CovP* and *AccP*. The key goal of DSPatch is to dynamically adapt prefetching for either higher coverage or higher accuracy depending on the DRAM bandwidth utilization. Using a simple 2-bit bandwidth utilization signal broadcast from the memory controller to all the cores, DSPatch selects either the *CovP* (when memory bandwidth utilization is low) or the *AccP* (when memory bandwidth utilization is high) bit-pattern to drive the prefetching.

Section 3.1 shows a high-level view of DSPatch. Section 3.2 discusses how DSPatch tracks the overall bandwidth utilization across all the cores. The subsequent sections describe the algorithm to modulate, learn and predict the spatial bit-patterns.

### 3.1 Overview

Figure 7 depicts the overall architecture of DSPatch. *Page Buffer* (*PB*) and *Signature Prediction Table* (*SPT*) are the two prime structures of DSPatch. Each *PB* entry tracks accesses in a 4KB physical page and accumulates L1 misses in the page's stored bit-pattern (step ①). The first access (step ②) to each 2KB segment in the 4KB physical page is eligible to trigger prefetches. The PC of this trigger access is stored in the *PB* entry and used to index into the *SPT*, which retrieves the two *CovP* and *AccP* bit-patterns and the measure of their goodness (step ③). Selection logic, detailed in Section 3.6, uses the memory bandwidth utilization measure to select a bit-pattern to generate prefetch candidates (step ④). The selected bit-pattern is anchored (i.e., rotated) to align to the trigger access offset before issuing prefetches. On eviction from the *PB* (step ⑤), for each trigger (per 2KB segment), the stored bit-pattern is first anchored (i.e., rotated) to trigger offset. Then, *SPT* is looked up using the stored trigger PC and the stored bit-patterns and the counters are updated as described in Section 3.6.

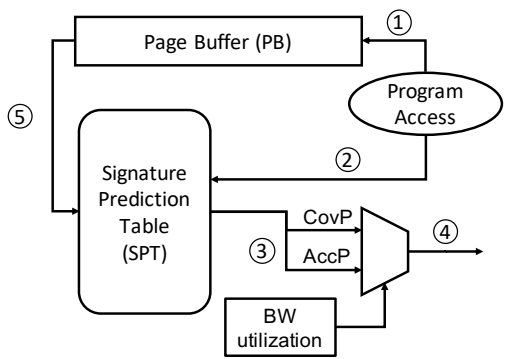

**Figure 7: DSPatch block diagram and overall organization**

### 3.2 Tracking Bandwidth Utilization

DSPatch tracks memory bandwidth utilization with a simple counter at the memory controller that counts the number of issued DRAM column access (CAS) commands in a time window of $(4 \times tRC)$ cycles (where $tRC$ is the minimum allowed time between two DRAM row activations). To include hysteresis in tracking, the counter is halved after every window. The number of channels and the width of each channel determines the peak DRAM bandwidth, as well as the peak possible number of CAS commands in each tRC window. We further bucket this counter into quartiles (25%, 50% and 75%) of peak bandwidth. Every tRC cycle, the value of the counter is compared to each of the three quartile thresholds, resulting in a two bit (2b) quantized value representing which quartile the current bandwidth utilization falls into (e.g., 3 indicates more than 75% bandwidth utilization, whereas 0 indicates less than 25% bandwidth utilization). This 2-bit quantized bandwidth utilization value is broadcast to all cores and is used as a representative of current memory bandwidth utilization in the DSPatch algorithm.

### 3.3 Anchored Spatial Bit-patterns

To maximize the possibility of exposing data access patterns, DSPatch uses a program access representation that is robust against reordering of accesses in the processor and the memory hierarchy. Figure 2 motivates the use of spatial bit-patterns anchored to the trigger (i.e., first) access to a memory region to capture all local and global deltas from the trigger. DSPatch uses such anchored bit-patterns to represent and predict program address access patterns to a given memory region. Our implementation of DSPatch employs a Page Buffer (*PB*) that tracks the 64 most-recently-accessed 4KB physical pages at the L2 cache level. Each *PB* entry stores a 64b bit-pattern that accumulates the L2 cache block addresses referenced by the program loads and stores in the page.

### 3.4 The Choice of Signature and Signature-Pattern Mapping

The choice of signature has a significant impact on the design of a prefetcher. The more information a signature encodes, the more prefetch filtering is achieved and hence the higher the expected accuracy. Prior bit-pattern prefetchers use the PC of the trigger access along with the offset in the page [73] or the actual page address [26]. They implicitly expect high prediction accuracy and hence just store and use the last occurring bit-pattern per signature without explicitly tracking accuracy. However, this comes at the cost of extra storage since the prefetcher needs to track a large enough set of frequently occurring signatures to achieve high coverage.

DSPatch uses just the PC of the trigger access to a physical page as the signature. It learns two modulated bit-patterns from the recently seen accesses to a physical page and stores the bit-patterns by associating them with a PC signature in the Signature-Pattern Table (*SPT*). Upon encountering the same signature, DSPatch looks up the SPT and selects a bit-pattern to generate prefetch candidates associated with that signature. DSPatch organizes the SPT as a 256-entry tagless direct-mapped structure. A simple folded-XOR hash of the PC is used to index into this structure. While this indexing can reduce storage requirements, there are associated trade-offs in accuracy and coverage. The use of only PC as a program signature can result in lower accuracy while aliasing of multiple PCs into a single entry can have an unpredictable impact if we only store and use the last occurring bit-pattern. Therefore, DSPatch uses simple mechanisms (with bitwise AND and PopCount operations) to track coverage and accuracy of stored bit-patterns (as we describe in Section 3.5). Crucially, DSPatch uses *two modulated* bit-patterns, one biased towards coverage (through OR operations) and the other biased towards accuracy (through AND operations). This allows DSPatch to simultaneously optimize for both coverage and accuracy (see Section 3.6). We describe these components of DSPatch in the subsequent sections.

### 3.5 Quantifying Accuracy and Coverage

Figure 8 depicts a simple scheme for quantifying the accuracy and coverage of bit-pattern predictions for a given physical page. PopCount of the predicted bit-pattern gives the prefetch count ($C_{pred}$), whereas the PopCount of the access bit-pattern generated by the program (called the *program bit-pattern*) gives the total number of accesses ($C_{real}$). Similarly, PopCount of the bitwise AND operation between the program bit-pattern and the predicted bit-pattern gives the accurate prefetch count ($C_{acc}$). **Prediction accuracy** is computed as the ratio $C_{acc}/C_{pred}$ whereas **prediction coverage** is computed as $C_{acc}/C_{real}$. Instead of computing the exact fractional value, we quantize our measure of accuracy and coverage into four quartiles via simple shift and compare operations.

### 3.6 Modulated Dual Bit-patterns: Coverage-biased and Accuracy-biased

A crucial goal of DSPatch is to have the ability to simultaneously optimize for prefetch coverage and accuracy based on memory

DSPatch: Dual Spatial Pattern Prefetcher

|  | Bit-pattern | PopCount |
|---|---|---|
| Program | 1011 0100 0011 1100 | 8 |
| Predicted | 1010 0110 0000 0001 | 5 |
| Bitwise-AND | 1010 0100 0000 0000 | 3 |

|  |  | <25% | 25-50% | 50-75% | >=75% |
|---|---|---|---|---|---|
| Prediction Accuracy | 3/5 |  |  | ✓ |  |
| Prediction Coverage | 3/8 |  | ✓ |  |  |

Figure 8: Prediction accuracy and coverage can be measured by simple bitwise AND and PopCount operations

bandwidth utilization. To this end, DSPatch stores *two modulated* bit-patterns per SPT entry, one is biased towards coverage (*CovP*) and the other is biased towards accuracy (*AccP*), as shown in Figure 9.

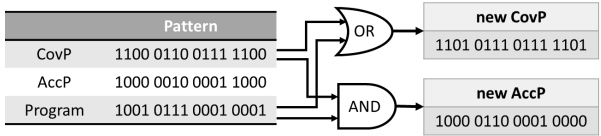

Figure 9: Two modulated spatial bit-patterns that can simultaneously optimize for both coverage and accuracy

**Coverage-biased Bit-pattern (*CovP*).** Since an anchored bit-pattern effectively captures all deltas from a trigger access, adding more deltas to increase predictions and coverage is a simple matter of setting the appropriate bits in the bit-pattern. This can be achieved via simple bitwise OR operations on the predicted bit-pattern with the program bit-pattern. However, since too many repeated ORs could eventually set all bits in a bit-pattern, we limit updates to at most three OR operations. DSPatch uses a 2b saturating counter named *OrCount* for each *CovP* to track the number of OR operations. *OrCount* is incremented every time the OR operation adds any bits to the predicted bit-pattern. DSPatch also employs a 2b saturating counter called $Measure_{CovP}$ to quantify the goodness of *CovP*. $Measure_{CovP}$ is incremented in two cases: (1) if the *CovP* prediction accuracy is less than a threshold value (called *AccThr*) or (2) if the prefetch coverage from *CovP* is less than a threshold value (called *CovThr*). Thus, a saturated $Measure_{CovP}$ value essentially indicates that prefetching with the *CovP* bit-pattern would either lack in prefetch accuracy or prefetch coverage, and hence *CovP* needs to be relearnt from scratch. DSPatch resets *CovP* to the current program bit-pattern when $Measure_{CovP}$ is saturated and either of the two following conditions are satisfied: (1) current memory bandwidth utilization is in the highest quartile or (2) prefetch coverage is less than 50%. We use the 50% quartile threshold value for both *AccThr* and *CovThr*.

**Accuracy-biased Bit-pattern (*AccP*).** The accuracy-biased bit-pattern requires retaining recurring bits in the bit-pattern, which can be achieved by an AND operation. Rather than recursive AND operations on *AccP*, on every update, *AccP* is replaced by a bitwise

AND operation of the program bit-pattern and the *CovP*. Similar to the $Measure_{CovP}$, DSPatch also uses a 2b saturating counter called $Measure_{AccP}$ to quantify the goodness of *AccP*. $Measure_{AccP}$ is incremented if *AccP* prediction accuracy is less than 50%, and is decremented otherwise. Thus, a saturated $Measure_{AccP}$ counter value essentially indicates that prefetching with the *AccP* bit-pattern would lack in prefetch accuracy. DSPatch uses $Measure_{AccP}$ to completely throttle down predictions when memory bandwidth utilization is high.

**Bit-pattern Selection for Prefetch Generation.** Figure 10 shows the algorithm DSPatch uses to choose between *CovP* and *AccP* for prefetch generation. When DRAM bandwidth utilization is in the highest quartile (75%), we select *AccP* for prefetching if $Measure_{AccP}$ is not saturated. When bandwidth utilization is in the second highest quartile (between 50% and 75%), we select *AccP* for prefetching if $Measure_{CovP}$ is saturated (indicating that *CovP* is inaccurate) and *CovP* otherwise. When bandwidth utilization is less than 50%, we simply select *CovP* for prefetching. To minimize any pollution effect when bandwidth utilization is less than 50%, we fill the prefetched blocks at low priority in the on-die L2 cache and LLC, if $Measure_{CovP}$ is saturated (indicating that *CovP* is inaccurate).

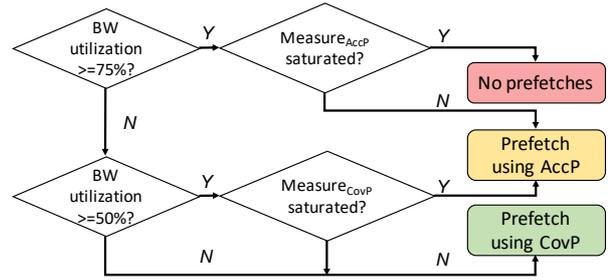

Figure 10: Selection of *CovP* versus *AccP* bit-patterns for prefetching based on DRAM bandwidth utilization and measures of the goodness of predictions

### 3.7 2KB (32b) vs 4KB (64b) Predictions and Multiple Triggers

Prior bit-pattern based prefetching proposals [37, 73] do not explicitly track accuracy and hence statically limit themselves to 2KB (32b) bit-patterns. Since DSPatch incorporates measures to track accuracy and to throttle its predictions, it can dynamically make predictions at both 2KB and 4KB memory region. Instead of using 64b bit-patterns for *CovP* and *AccP*, we split them into two 32b bit-patterns. The 2b $Measure_{CovP}$ and $Measure_{AccP}$ counters track 2KB (32b) bit-patterns and the prefetch generation is done per 2KB (32b) segment of a 4KB page. Splitting a 64b bit-pattern into two 32b bit-patterns also enables a further benefit for DSPatch: *two prefetch triggers per 4KB page (one per 2KB segment)*. The first (trigger) access to each 2KB segment in the 4KB page can attempt to trigger prefetches. The trigger to the first 2KB segment is allowed to predict both 32b bit-patterns (i.e., for the full 4KB page) while the trigger to the second 2KB segment is only allowed to predict a single 32b bit-pattern (for the 2KB region relative to the trigger).

## 3.8 Compressing Bit-patterns to Further Reduce Storage Requirements

DSPatch uses one final optimization to further reduce the storage overhead of the bit-patterns. We see that deltas +1 and -1 are the two most frequently occurring deltas in programs. As shown in Figure 11(a), these two deltas together appear more than 50% of the time on average. Therefore, instead of storing bit-patterns with each bit representing a 64B cacheline, we store a compressed bit-pattern where each bit represents *two adjacent* 64B cachelines. We call this compression technique 128B-granularity compression and the resultant compressed bit-pattern 128B-granularity bit-pattern. 128B-granularity compression halves DSPatch's pattern storage requirements. While this compression technique could theoretically have up-to 50% inaccuracy in predictions, we observe *less than one misprediction for every five cacheline predictions* (20% inaccuracy). Figure 11(b) shows the distribution of misprediction rate caused by 128B-granularity compression across all of our evaluated single-threaded workloads. As we can see from the figure, 128B-granularity compression incurs no mispredictions 42% of the time, across all workloads. This also means, 128B-granularity compression is able to represent the exact bit-pattern by consuming only half of the storage in 42% of the time. In fact, for 70% of the time, the misprediction rate caused by 128B-granularity compression is lower than 25%.

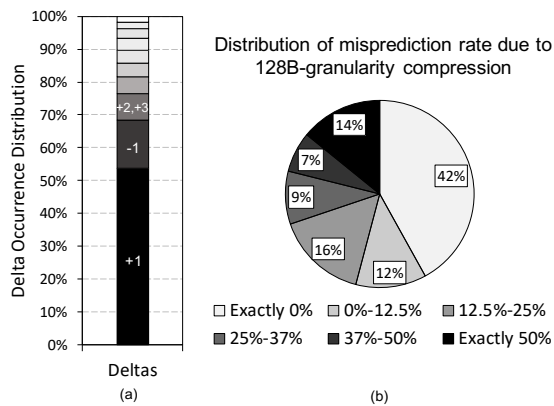

Figure 11: (a) +1 and -1 are the two most frequently occurring deltas, occurring more than 60% of time across all workloads. (b) 128B-granularity compression induces no mispredictions 42% of the time.

## 3.9 Storage Requirements

Table 1 shows that DSPatch requires only 3.6KB of storage for the configuration we evaluate in Section 5.

| Structure | Field (#bits in each entry) | Entries | #Bits |
|---|---|---|---|
| PB | Page number (36) + Bit-pattern (64) + 2x[PC (8) + Offset (6)] = 158 bits | 64 | 10112 |
| SPT | CovP (32) + 2*$Measure_{CovP}$ (2) + 2*ORCount (2) + AccP (32) + 2*$Measure_{AccP}$ (2) = 76 bits | 256 | 19456 |
| Total | | | 3.6 KB |

Table 1: DSPatch storage overhead

## 4 METHODOLOGY

We evaluate DSPatch using an in-house cycle accurate simulator that models dynamically-scheduled x86 cores clocked at 4 GHz. The core micro-architectural parameters are taken from the latest Intel Skylake processor [1] and are listed in Table 2. Single-thread (ST) simulations use a 2MB LLC and a single DDR4-2133 channel, while multi-programmed (MP) simulations use an 8MB LLC shared across four cores and two DDR4-2133 channels. Therefore, both ST and MP configurations provide the same LLC capacity per core but MP configurations have half the memory bandwidth per core (8.5GBps) of ST (17GBps).

| Core | 4 cores, 4-wide OoO, 224-entry ROB, 80-entry load buffer |
|---|---|
| L1 cache | Private, 32KB, 64B line, 8 way, LRU, 16 MSHRs, 5-cycle round-trip latency |
| L2 cache | Private, 256KB, 64B line, 8 way, LRU, 32 MSHRs, 8-cycle round-trip latency |
| LLC | ST: 2MB; MP: shared 8MB, 64B line, 16 way, Prefetch aware dead-block predictor similar to [39], 32 MSHRs per LLC Bank, 30-cycle round-trip latency |
| Main Memory | ST: Single channel; MP: Dual channel DDR4-2133MHz, 2 ranks/channel, 8 banks/rank, 64b data bus width per channel, 2KB row buffer/bank, tCL=15ns, tRCD=15ns, tRP=15ns, tRAS=39ns |
| L1D prefetch | PC-based stride prefetcher [38], tracks 64 PCs |

Table 2: Simulation parameters

### 4.1 Prefetchers

Our baseline configuration has a PC-based stride prefetcher in the L1 cache. We evaluate prior prefetching proposals SMS [73], BOP [62] and SPP [54], as well as DSPatch, as the L2 prefetcher. Each L2 prefetcher is trained on L1 misses (both demand and prefetch misses from L1) and fills prefetched lines into the L2 cache and the LLC. We fine tune each prefetcher individually in our simulation environment to produce the best possible result. Table 3 shows the prefetcher configurations. We also evaluate the AMPM prefetcher [43] but do not show its results as it under-performs all other prefetchers in single-thread simulations.

| BOP [62] | 256-entry RR, MaxRound=100, MaxScore=31, BadScore=1, Degree=2 (for ST), 1 (for MT) | 1.3 KB |
|---|---|---|
| SMS [73] | 2KB page region, 64-entry AT, 32-entry FT, 16K-entry PHT | 88 KB |
| SPP [54] | 256-entry ST, 512-entry PT, 8-entry GHR, 12b compressed delta path, 10b feedback | 6.2KB |

Table 3: Parameters of each evaluated prefetcher

### 4.2 Workloads

We evaluate a diverse set of 75 workloads, *including all benchmarks from the SPEC CPU2017 [17] and the SPEC CPU2006 [16] suites*. These workloads span various types of real-world applications, and we categorize them into 9 classes. Table 4 shows the workload classes along with example workloads from each class. We present the average performance of each of these classes as well as the geometric mean performance across all 75 workloads.

We use both homogeneous and heterogeneous workload mixes to simulate a multi-programmed system. To construct the homogeneous workload mixes, we select each of the 42 high-MPKI workloads from our full workload set and run four copies of it, one in each core of the simulator. To construct the heterogeneous workload mixes, we randomly select four workloads from the 42 high-MPKI workloads and generate 75 heterogeneous workload mixes.

DSPatch: Dual Spatial Pattern Prefetcher

| Class | Example Workloads |
|---|---|
| Client | 7-zip compression and decompression [2], vp9-encoding/decoding [23] |
| Server | TPC-C [22], SPECjbb 2015 [19], SPECjEnterprise2010 [20], Spark pagerank [5] |
| HPC | linpack [9], NAS Parallel Benchmarks [13], PARSEC [14], SPEC-ACCEL [15], SPEC MPI [18] |
| FSPEC06 | All benchmarks. e.g., sphinx3, soplex, GemsFDTD |
| ISPEC06 | All benchmarks. e.g., gcc, mcf, omnetpp |
| FSPEC17 | All benchmarks. e.g., namd, povray, lbm |
| ISPEC17 | All benchmarks. e.g., omnetpp, xalancbmk, leela |
| Cloud | Bigbench [6], Cassandra [3], Hadoop-hbase, kmeans, streaming [4] |
| SYSMark | SYSmark-excel, photoshop, word, sketchup [21] |

Table 4: Evaluated workload categories

## 5 EVALUATION

### 5.1 Single-thread Performance

Figure 12 shows the performance comparison of prior prefetchers and DSPatch, both as a standalone prefetcher and as an adjunct prefetcher to SPP. We make three major observations from Figure 12. First, as a standalone prefetcher, DSPatch outperforms SMS by 3% on average across all workloads while requiring only less than $1/20^{th}$ of the storage requirements of SMS. The use of dual modulated bit-patterns and multiple triggers helps DSPatch outperform the traditional bit-pattern prefetching employed by SMS. Second, DSPatch performs 1% better than SPP on average while requiring only less than $2/3^{rd}$ of the storage of SPP. DSPatch's gains over SPP mainly come from the SYSmark, Cloud and ISPEC workload categories. All other workload categories showcase the benefits of the fine-grained delta-based prefetching paradigm employed by SPP. Third, the combination of fine-grained delta-based prefetching of SPP and dual modulated spatial bit-pattern-based prefetching of DSPatch achieves 6% performance improvement over standalone SPP on average, *outperforming every standalone prefetcher in all workload categories*. This clearly makes the case for using DSPatch as a lightweight adjunct prefetcher to SPP so that we can extract the benefits of both prefetching paradigms.

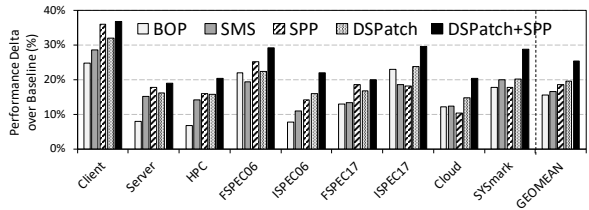

Figure 12: Single-thread performance results

Figure 13 shows the performance line graph for the set of 42 memory-intensive single-thread workloads. On average, the combined DSPatch+SPP prefetcher outperforms the standalone SPP by 9%. DSPatch+SPP performs worse than SMS in only one `TPC-C` workload, which has very large code footprint. With more than 4000 trigger PCs per kilo instructions, SMS benefits from its large signature storage capability of 16K entries. DSPatch+SPP outperforms standalone SPP by 26% in NPB, by 20% in BigBench and by 16% in SYSMark-excel and mcf (ISPEC06) workloads.

Figure 14 shows the performance of BOP and a 256-entry SMS (iso-storage with DSPatch) as adjunct prefetchers to SPP. We make two observations from Figure 14. First, as an adjunct prefetcher to SPP, DSPatch provides much higher performance than BOP and SMS with similar storage requirements. Second, DSPatch+SPP outperforms BOP+SPP by 2.1%, mainly because DSPatch+SPP has higher prefetch coverage than BOP+SPP (60% vs 55%).

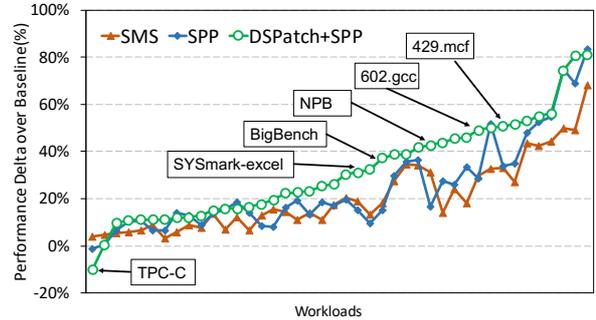

Figure 13: Performance line graph for 42 memory-intensive single-thread workloads.

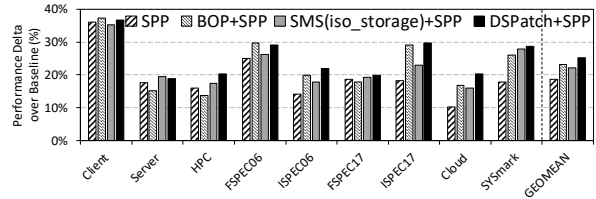

Figure 14: Performance of BOP, 256 entry SMS and DSPatch as adjunct prefetchers to SPP

For comprehensiveness, we also evaluate DSPatch in conjunction with SPP and BOP together. DSPatch improves average performance by 2.6% on top of the SPP+BOP combination prefetcher on average (not shown here). This clearly indicates non-overlapping coverage opportunities between BOP and DSPatch, encouraging further optimization and research in this direction.

### 5.2 Performance Scaling with Memory Bandwidth

Figure 15 shows how the performance improvements of different prefetchers scale as we scale the DRAM bandwidth from a single channel DDR4-1600 (with 12.5GBps bandwidth) to dual channel DDR4-2400 (with 38 GBps bandwidth).

Two key takeaways emerge from the data. First, the performance of DSPatch+SPP scales well with increasing memory bandwidth, growing from 6% over SPP to 10% when the memory bandwidth is doubled, when going from the single channel DDR4-2133 system to the dual channel DDR4-2133 system. Second, as an adjunct prefetcher to SPP, the performance gap between eBOP+SPP and DSPatch+SPP increases with increase in memory bandwidth headroom, growing from 2.1% in the single channel DDR4-2133 system to 5% in the dual channel DDR4-2400 system. We conclude that, DSPatch, via its fundamental design choices, is best suited to extract higher performance from higher memory bandwidth.

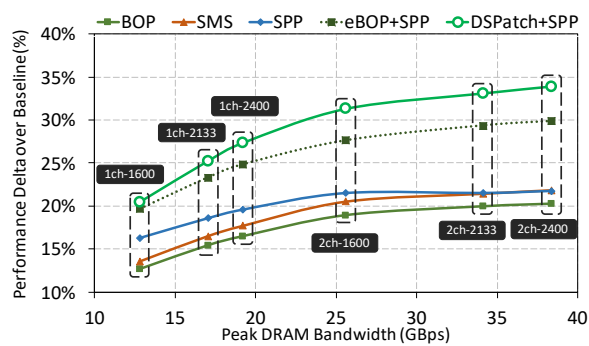

**Figure 15: Performance scaling with DRAM bandwidth**

### 5.3 Impact On Coverage And Accuracy

Figure 16 quantifies the coverage and misprediction rates of the evaluated prefetchers. On average, DSPatch+SPP has 15% higher coverage than the standalone SPP prefetcher. This comes at a 6.5% increase in the rate of mispredictions. Since DSPatch uses dual modulated patterns simultaneously optimized for both coverage and accuracy, we achieve a 2:1 ratio in the impact on coverage and accuracy: a 2% increase in coverage comes at only a 1% increase in mispredictions.

### 5.4 Multi-programmed Performance

Multiple cores competing for the DRAM bandwidth resource reduces the headroom for prefetchers to boost coverage and performance. The use of the accuracy-biased bit-pattern in DSPatch plays a crucial role in such scenarios to generate highly accurate prefetches to make the best use of the scarce DRAM bandwidth.

Figure 17 shows the performance improvement of all prefetchers on 42 homogeneous workload mixes. We make two observations. First, as an adjunct prefetcher to SPP, DSPatch improves performance by 5.9% over the standalone SPP. Second, even though SMS outperforms the standalone SPP by 3.2% and 2.5% in the Cloud and SYSmark workload categories, DSPatch+SPP outperforms SMS by 4% and 7.3% in these workload categories.

Figure 18 compares the performance improvement of all prefetchers on homogeneous and heterogeneous workload mixes for two different DRAM bandwidth configurations: dual channel DDR4 at

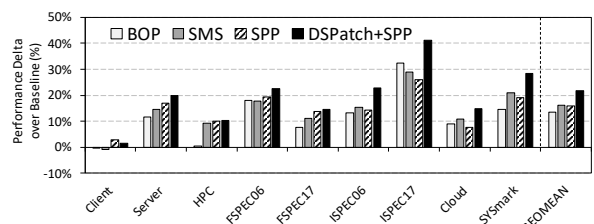

**Figure 17: Multi-programmed performance results across 42 homogeneous workload mixes**

2133 MHz and 2400 MHz. We make two observations. First, in the baseline system with two DDR4-2133 channels, DSPatch improves average performance by 7.4% over the standalone SPP across all heterogeneous mixes. Second, increasing the DRAM channel frequency by 12.5% (i.e., from 2133 MHz to 2400MHz) also increases performance by 7.7% and DSPatch+SPP outperforms the standalone SPP by 15.1%. We conclude that DSPatch maintains the ability to scale up performance based on available memory bandwidth even when memory bandwidth is a scarce resource.

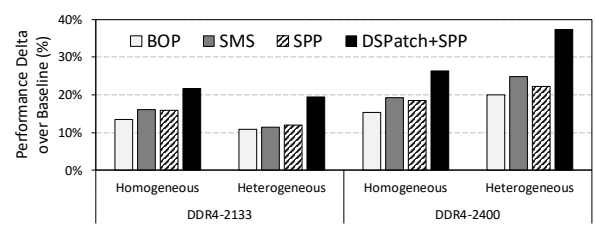

**Figure 18: Performance improvement of prefetchers on homogeneous and heterogeneous multi-programmed workload mixes for two different DRAM bandwidths**

### 5.5 Contribution Of Accuracy-biased Patterns

We study the performance impact and contribution of the accuracy-biased predictions in the DSPatch design. In DSPatch, accuracy-biased bit-patterns provide highly accurate predictions when memory bandwidth utilization is close to peak (greater than 75%). Figure 19 shows the full-blown DSPatch's performance improvement along with two other configurations of DSPatch that never use *AccP* for prediction: (1) one DSPatch configuration (named AlwaysCovP) always uses *CovP* for prediction, even when DRAM bandwidth utilization is high and (2) another DSPatch configuration (named `ModCovP`) dynamically throttles down *CovP* to restrict aggressiveness when DRAM bandwidth utilization is high. We make two observations from the figure. First, always using *CovP* for prediction irrespective of the memory bandwidth utilization significantly hurts performance. `AlwaysCovP` loses 4.5% performance over the full-blown DSPatch. Second, throttling down *CovP* prediction at high DRAM bandwidth utilization scenarios alone does not solve the problem: `ModCovP` loses 1.4% performance over the full-blown DSPatch. We conclude that statically selecting any single type of bit-pattern is sub-optimal in performance and we need two modulated bit-patterns to enable a dynamic selection of the appropriate bit-pattern at run-time.

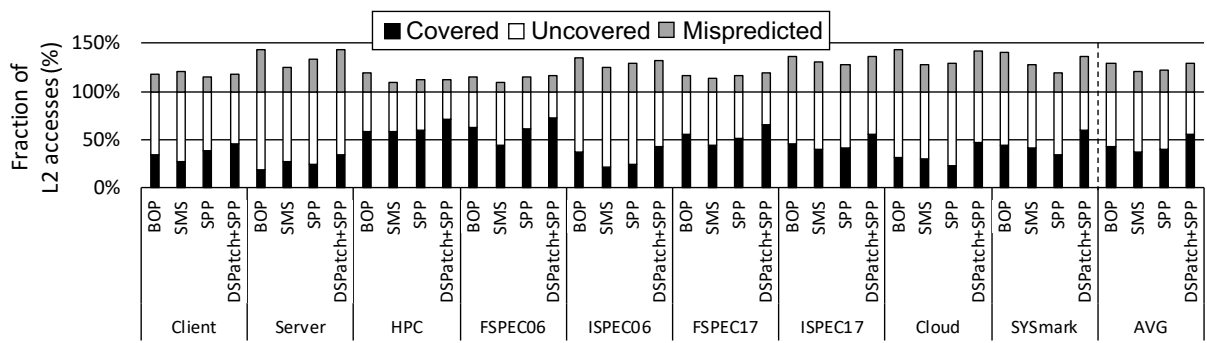

**Figure 16: Coverage and mispredictions of different prefetchers**

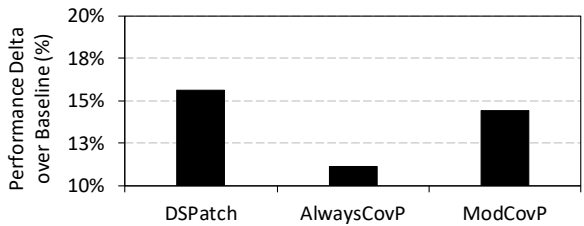

**Figure 19: Performance improvement of the full-blown DSPatch versus two other DSPatch variants that do not use the accuracy-biased bit-pattern. The y-axis starts at 10%.**

## 6 RELATED WORK

Prefetching is an extensively studied approach to hide high memory latency with a large body of work over decades covering a wide range of algorithms and implementations. To our knowledge, this is the first work to use two spatial bit-patterns and a memory-bandwidth-driven dynamic selection of bit-patterns to achieve better scalability in performance with scaling in memory bandwidth.

We divide past prefetching solutions into three major categories: pre-computation, temporal, and non-temporal prefetchers. We compare DSPatch with each of these categories, as well as prior prefetch-throttling mechanisms and highlight how DSPatch fundamentally differs from them.

**Pre-computation Prefetchers.** One flavor of prefetching relies on pre-computation to hide latency. Examples include runahead execution [32, 41, 51, 63–65] and helper thread prefetching [28, 30, 60, 76, 80] proposals. The use of pre-computation makes these proposals highly accurate with the ability to provide coverage even when no patterns exist in address accesses. However, pre-computation prefetchers are higher in complexity compared to light-weight hardware prefetchers that capture access patterns. DSPatch, being a traditional prefetching proposal, differs completely from pre-computation prefetchers and predicts future accesses only by learning patterns in past accesses.

**Temporal Prefetchers.** Temporal prefetchers including STeMS [77], ISB [45] and the Domino prefetcher [25] are built on the Markov prefetching [49] model by tracking the temporal order of full cache-line address accesses rather than address deltas or cache-line offsets in spatial regions. While tracking repeating patterns of full cacheline addresses can be quite accurate, it has multi-megabyte storage requirements, which necessitates storing meta-data in memory. DSPatch requires only 3.6 KB, which can easily fit inside a core.

**Non-temporal Prefetchers.** Prefetchers that predict deltas or bit-patterns in a spatial region (like a 2KB or 4KB page) have significantly lower storage requirements and generally lower complexity than pre-computation or temporal prefetchers. Stream [29, 50] and stride [38] prefetchers capture simple repeating deltas. More recently, prefetching proposals that can capture more complex delta patterns like a repeating series of deltas have emerged. We further categorize these prefetchers into two broad groups.

**(1) Delta-based Prefetchers.** Delta-based prefetchers like VLDP [72] and SPP [54] use a history of address deltas (inspired by the TAGE [71] branch predictor) to predict future deltas. As we discussed and evaluated, SPP uses its prefetch confidence values to recursively prefetch further ahead to improve timeliness. BOP [62] uses a set of global deltas to capture a repeating series of smaller deltas. We have already comprehensively compared DSPatch to both SPP and BOP proposals in this work. Another prior proposal, Kill-the-PC [55] co-designs both the prefetching and the cache replacement policy to be aware of each other. However, the prefetching component of KPC (called KPC-P) is identical to SPP. Our evaluation of the prefetching component of KPC showed no significant improvement over SPP.

**(2) Bit-pattern-based Prefetchers.** Bit-pattern-based prefetchers, exemplified by SMS [73], use the PC as part of their signature to predict bit-patterns. These prefetchers have higher storage requirements (of the order of many tens of KB) than delta-based prefetchers. Rotated bit-patterns were first used by Ferdman et al. [37] to eliminate the cacheline offset in the spatial region (2KB) from the signature and reduce storage requirements to around 40KB. DSPatch further reduces storage by compressing the bit-pattern where each bit represents one 128B block, instead of a single 64B cacheline. A recent work, Bingo [26], extends bit-pattern-based prefetching to use both long and short history events (again inspired by the TAGE [71] branch predictor). In addition to the offset in the region along with the PC as part of the signature (a short event in their terminology), Bingo also supports a long event signature utilizing the full cacheline address. Bingo fuses these signatures into the same prediction table, enable multiple predictions from a single entry for higher coverage than SMS. However, Bingo still consumes over 100KB of area. DSPatch significantly simplifies bit-pattern prefetching using a mere 3.6KB of storage with anchored bit-patterns along with mechanisms to track and boost coverage and accuracy for higher performance. DSPatch's design choices fundamentally enable good performance scaling with higher memory bandwidth as well.

**Prefetch-throttling Mechanisms.** Prefetcher throttling mechanisms play a crucial role in any aggressive prefetcher design. Multiple prior proposals [31, 34–36, 42, 56, 57, 67, 69, 70, 74, 79] take prefetching metrics like coverage, accuracy and bandwidth consumption into consideration to selectively throttle or drop prefetch requests in an attempt to reduce prefetcher-induced pollution in cache capacity as well as in available memory bandwidth. DSPatch also has inherently simple mechanisms to track prefetch accuracy and coverage, along with the ability to scale performance with increase in memory bandwidth. Even so, prior prefetch-throttling proposals can be orthogonally applied to DSPatch as well to further adjust its prefetch aggressiveness.

## 7 SUMMARY

We introduce DSPatch, a new spatial bit-pattern prefetcher that uses memory bandwidth utilization inherently in its algorithm to adjust prefetch aggressiveness and provide better scaling in performance improvement with increase in memory bandwidth. DSPatch exploits two key ideas. First, it learns *two* spatial bit-patterns to generate prefetches in a given memory region (i.e., a physical page) by using simple logical OR and AND operations. One bit-pattern is biased towards coverage and the other bit-pattern is biased towards accuracy. Second, DSPatch dynamically selects any one bit-pattern to generate prefetches at run-time based on the memory bandwidth utilization and the coverage and accuracy of each bit-pattern. These two ideas in unison help DSPatch to achieve better scaling in performance with increase in memory bandwidth than state-of-the-art prefetchers. Our evaluations show that, using only 3.6 KB of hardware storage, DSPatch improves performance by 6%, on average across 75 single-thread workloads, over an aggressive baseline with a PC-based stride prefetcher at the L1 cache and the SPP prefetcher at the L2 cache. DSPatch's performance improvement grows from 6% to 10% when DRAM bandwidth is doubled. As memory bandwidth continues to increase with improvements in DRAM architecture and packaging, we believe that the next-generation processors will significantly benefit from DSPatch's ability to extract higher performance in the presence of higher memory bandwidth.

## ACKNOWLEDGMENTS

We thank the anonymous reviewers for their useful feedback. We also thank all the members of Intel Processor Architecture Research Lab, especially Shankar Balachandran for his immense help and constructive feedback.

## REFERENCES

[1] "6th Generation Intel® Processor Family," https://www.intel.com/content/www/us/en/processors/core/desktop-6th-gen-core-family-spec-update.html.
[2] "7-Zip," https://www.7-zip.org/.
[3] "Apache Cassandra," https://cassandra.apache.org/.
[4] "Apache Hadoop," https://hadoop.apache.org/.
[5] "Apache Spark$^{TM}$," https://www.cloudera.com/products/open-source/apache-hadoop/apache-spark.html.
[6] "BigBench," https://blog.cloudera.com/blog/2014/11/bigbench-toward-an-industry-standard-benchmark-for-big-data-analytics/.
[7] "HBM Specification," https://www.amd.com/Documents/High-Bandwidth-Memory-HBM.pdf.
[8] "HMC Specification v2.1," http://www.hybridmemorycube.org/files/SiteDownloads/HMC-30G-VSR_HMCC_Specification_Rev2.1_20151105.pdf.
[9] "HP-LINPACK," https://www.netlib.org/benchmark/hpl/.
[10] "JEDEC-DDR4," https://www.jedec.org/sites/default/files/docs/JESD79-4.pdf.
[11] "JEDEC-GDDR5," https://www.jedec.org/category/keywords/gddr5.
[12] "LPDDR4 Specification," https://www.jedec.org/sites/default/files/docs/JESD209-4.pdf.
[13] "NAS Parallel Benchmark," https://github.com/benchmark-subsetting/NPB3.0-omp-C.
[14] "PARSEC," http://parsec.cs.princeton.edu/.
[15] "SPEC ACCEL©," https://www.spec.org/accel/.
[16] "SPEC CPU 2006," https://www.spec.org/cpu2006/.
[17] "SPEC CPU 2017," https://www.spec.org/cpu2017/.
[18] "SPEC MPI© 2007," https://www.spec.org/mpi2007/.
[19] "SPECjbb© 2015," https://www.spec.org/jbb2015/.
[20] "SPECjEnterprise© 2010," https://www.spec.org/jEnterprise2010/.
[21] "SYSmark 2014 ver 1.5," https://bapco.com/products/sysmark-2014/.
[22] "TPC-C," http://www.tpc.org/tpcc/detail.asp.
[23] "VP9 Encoding," https://trac.ffmpeg.org/wiki/Encode/VP9.
[24] J. Ahn, S. Hong, S. Yoo, O. Mutlu, and K. Choi, "A Scalable Processing-in-memory Accelerator for Parallel Graph Processing," in *ISCA*, 2015.
[25] M. Bakhshalipour, P. Lotfi-Kamran, and H. Sarbazi-Azad, "Domino Temporal Data Prefetcher," in *HPCA*, 2018.
[26] M. Bakhshalipour, M. Shakerinava, P. Lotfi-Kamran, and H. Sarbazi-Azad, "Bingo spatial data prefetcher," in *HPCA*, 2019.
[27] K. K. Chang, A. Kashyap, H. Hassan, S. Ghose, K. Hsieh, D. Lee, T. Li, G. Pekhimenko, S. Khan, and O. Mutlu, "Understanding Latency Variation in Modern DRAM Chips: Experimental Characterization, Analysis, and Optimization," in *SIGMETRICS*, 2016.
[28] R. S. Chappell, J. Stark, S. P. Kim, S. K. Reinhardt, and Y. N. Patt, "Simultaneous Subordinate Microthreading (SSMT)," in *ISCA*, 1999.
[29] T.-F. Chen and J.-L. Baer, "Effective hardware-based data prefetching for high-performance processors," in *IEEE TC*, 1995.
[30] J. D. Collins, D. M. Tullsen, H. Wang, and J. P. Shen, "Dynamic Speculative Precomputation," in *MICRO*, 2001.
[31] F. Dahlgren, M. Dubois, and P. Stenström, "Sequential Hardware Prefetching in Shared-Memory Multiprocessors," in *IEEE TPDS*, 1995.
[32] J. Dundas and T. Mudge, "Improving Data Cache Performance by Pre-executing Instructions Under a Cache Miss," in *ICS*, 1997.
[33] E. Ebrahimi, C. J. Lee, O. Mutlu, and Y. N. Patt, "Fairness via Source Throttling: A Configurable and High-performance Fairness Substrate for Multi-core Memory Systems," in *ASPLOS*, 2010.
[34] E. Ebrahimi, C. J. Lee, O. Mutlu, and Y. N. Patt, "Prefetch-aware Shared Resource Management for Multi-core Systems," in *ISCA*, 2011.
[35] E. Ebrahimi, O. Mutlu, C. J. Lee, and Y. N. Patt, "Coordinated control of multiple prefetchers in multi-core systems," in *MICRO*, 2009.
[36] E. Ebrahimi, O. Mutlu, and Y. N. Patt, "Techniques for bandwidth-efficient prefetching of linked data structures in hybrid prefetching systems," in *HPCA*, 2009.
[37] M. Ferdman, S. Somogyi, and B. Falsafi, "Spatial memory streaming with rotated patterns," in *In 1st JILP Data Prefetching Championship*, 2009.
[38] J. W. C. Fu, J. H. Patel, and B. L. Janssens, "Stride Directed Prefetching in Scalar Processors," in *MICRO*, 1992.
[39] J. Gaur, M. Chaudhuri, and S. Subramoney, "Bypass and insertion algorithms for exclusive last-level caches," in *ISCA*, 2011.
[40] S. Ghose, T. Li, N. Hajinazar, D. Senol Cali, and O. Mutlu, "Demystifying Complex Workload-DRAM Interactions: An Experimental Study," in *SIGMETRICS*, 2019.
[41] M. Hashemi, O. Mutlu, and Y. N. Patt, "Continuous runahead: Transparent hardware acceleration for memory intensive workloads," in *MICRO*, 2016.
[42] I. Hur and C. Lin, "Memory Prefetching Using Adaptive Stream Detection," in *MICRO*, 2006.
[43] Y. Ishii, M. Inaba, and K. Hiraki, "Access map pattern matching for data cache prefetch," in *ISC*, 2009.
[44] Y. Ishii, M. Inaba, and K. Hiraki, "Unified Memory Optimizing Architecture: Memory Subsystem Control with a Unified Predictor," in *ICS*, 2012.
[45] A. Jain and C. Lin, "Linearizing irregular memory accesses for improved correlated prefetching," in *MICRO*, 2013.
[46] A. Jain and C. Lin, "Rethinking belady's algorithm to accommodate prefetching," in *ISCA*, 2018.
[47] V. Janapa Reddi, B. C. Lee, T. Chilimbi, and K. Vaid, "Web Search Using Mobile Cores: Quantifying and Mitigating the Price of Efficiency," in *ISCA*, 2010.
[48] D. A. Jiménez, "Dead block replacement and bypass with a sampling predictor," in *JWAC*, 2010.
[49] D. Joseph and D. Grunwald, "Prefetching using Markov predictors," in *ISCA*, 1997.
[50] N. P. Jouppi, "Improving Direct-mapped Cache Performance by the Addition of a Small Fully-associative Cache and Prefetch Buffers," in *ISCA*, 1990.
[51] D. Kadjo, J. Kim, P. Sharma, R. Panda, P. Gratz, and D. Jimenez, "B-fetch: Branch prediction directed prefetching for chip-multiprocessors," in *MICRO*, 2014.
[52] S. Kanev, J. P. Darago, K. Hazelwood, P. Ranganathan, T. Moseley, G.-Y. Wei, and D. Brooks, "Profiling a Warehouse-scale Computer," in *ISCA*, 2015.
[53] S. M. Khan, Y. Tian, and D. A. Jimenez, "Sampling dead block prediction for last-level caches," in *MICRO*, 2010.
[54] J. Kim, S. H. Pugsley, P. V. Gratz, A. Reddy, C. Wilkerson, and Z. Chishti, "Path confidence based lookahead prefetching," in *MICRO*, 2016.
[55] J. Kim, E. Teran, P. V. Gratz, D. A. Jiménez, S. H. Pugsley, and C. Wilkerson, "Kill the Program Counter: Reconstructing Program Behavior in the Processor Cache Hierarchy," in *ASPLOS*, 2017.
[56] C. J. Lee, O. Mutlu, V. Narasiman, and Y. N. Patt, "Prefetch-aware DRAM controllers," in *MICRO*, 2008.
[57] C. J. Lee, V. Narasiman, O. Mutlu, and Y. N. Patt, "Improving Memory Bank-level Parallelism in the Presence of Prefetching," in *MICRO*, 2009.
[58] D. Lee, S. Ghose, G. Pekhimenko, S. Khan, and O. Mutlu, "Simultaneous multi-layer access: Improving 3D-stacked memory bandwidth at low cost," in *TACO*, 2016.
[59] D. Lee, Y. Kim, G. Pekhimenko, S. Khan, V. Seshadri, K. Chang, and O. Mutlu, "Adaptive-latency DRAM: Optimizing DRAM timing for the common-case," in

*HPCA*, 2015.
[60] C.-K. Luk, "Tolerating Memory Latency Through Software-controlled Pre-execution in Simultaneous Multithreading Processors," in *ISCA*, 2001.
[61] K. T. Malladi, F. A. Nothaft, K. Periyathambi, B. C. Lee, C. Kozyrakis, and M. Horowitz, "Towards energy-proportional datacenter memory with mobile DRAM," in *ISCA*, 2012.
[62] P. Michaud, "Best-offset hardware prefetching," in *HPCA*, 2016.
[63] O. Mutlu, H. Kim, and Y. N. Patt, "Techniques for efficient processing in runahead execution engines," in *ISCA*, 2005.
[64] O. Mutlu, H. Kim, and Y. N. Patt, "Efficient runahead execution: Power-efficient memory latency tolerance," in *IEEE Micro*, 2006.
[65] O. Mutlu, J. Stark, C. Wilkerson, and Y. N. Patt, "Runahead execution: An alternative to very large instruction windows for out-of-order processors," in *HPCA*, 2003.
[66] O. Mutlu, J. Stark, C. Wilkerson, and Y. N. Patt, "Runahead Execution: An Effective Alternative to Large Instruction Windows," in *IEEE Micro*, 2003.
[67] B. Panda and S. Balachandran, "Expert Prefetch Prediction: An Expert Predicting the Usefulness of Hardware Prefetchers," in *IEEE CAL*, 2016.
[68] D. Pandiyan, S.-Y. Lee, and C.-J. Wu, "Performance, Energy Characterizations and Architectural Implications of An Emerging Mobile Platform Benchmark Suite-MobileBench," in *IISWC*, 2013.
[69] S. H. Pugsley, Z. Chishti, C. Wilkerson, P.-f. Chuang, R. L. Scott, A. Jaleel, S.-L. Lu, K. Chow, and R. Balasubramonian, "Sandbox prefetching: Safe run-time evaluation of aggressive prefetchers," in *HPCA*, 2014.
[70] V. Seshadri, S. Yedkar, H. Xin, O. Mutlu, P. B. Gibbons, M. A. Kozuch, and T. C. Mowry, "Mitigating prefetcher-caused pollution using informed caching policies for prefetched blocks," in *TACO*, 2015.
[71] A. Seznec, "A new case for the TAGE branch predictor," in *MICRO*, 2011.
[72] M. Shevgoor, S. Koladiya, R. Balasubramonian, C. Wilkerson, S. H. Pugsley, and Z. Chishti, "Efficiently prefetching complex address patterns," in *MICRO*, 2015.
[73] S. Somogyi, T. F. Wenisch, A. Ailamaki, B. Falsafi, and A. Moshovos, "Spatial memory streaming," in *ISCA*, 2006.
[74] S. Srinath, O. Mutlu, H. Kim, and Y. N. Patt, "Feedback directed prefetching: Improving the performance and bandwidth-efficiency of hardware prefetchers," in *HPCA*, 2007.
[75] L. Tang, J. Mars, N. Vachharajani, R. Hundt, and M. L. Soffa, "The Impact of Memory Subsystem Resource Sharing on Datacenter Applications," in *ISCA*, 2011.
[76] P. H. Wang, J. D. Collins, H. Wang, D. Kim, B. Greene, K.-M. Chan, A. B. Yunus, T. Sych, S. F. Moore, and J. P. Shen, "Helper Threads via Virtual Multithreading on an Experimental Itanium®2 Processor-based Platform," in *ASPLOS*, 2004.
[77] T. F. Wenisch, M. Ferdman, A. Ailamaki, B. Falsafi, and A. Moshovos, "Practical off-chip meta-data for temporal memory streaming," in *HPCA*, 2009.
[78] C.-J. Wu, A. Jaleel, W. Hasenplaugh, M. Martonosi, S. C. Steely Jr, and J. Emer, "SHiP: Signature-based hit predictor for high performance caching," in *MICRO*, 2011.
[79] C.-J. Wu, A. Jaleel, M. Martonosi, S. C. Steely Jr, and J. Emer, "PACMan: prefetch-aware cache management for high performance caching," in *MICRO*, 2011.
[80] C. Zilles and G. Sohi, "Execution-based Prediction Using Speculative Slices," in *ISCA*, 2001.

## APPENDIX: EFFECT OF INACCURATE PREFETCHES ON CACHE POLLUTION

To study the impact of pollution caused by inaccurate prefetches in the last level cache (LLC), we examine all load, store and prefetch requests (where prefetches are generated by an aggressive but fairly inaccurate streaming prefetcher [29]) for all of our simulated workloads, using the methodology we describe in Section 4. We capture request type (load, store or prefetch), address of request, and for every level of cache that a request fills into, the victim address it evicts from that level. Inaccurate prefetches are those prefetches that do not see use by a demand request (load or store) before their eviction from on-die caches. We use the LLC victim addresses evicted by inaccurate prefetches to quantify the cache pollution caused by the prefetcher. We categorize the LLC victim addresses into three classes. Figure 20 illustrates the breakdown of these classes across all of our 75 single-threaded workloads for three different LLC sizes.

- **NoReuse**: LLC victim addresses that see no use by any demand within 10 million instructions of their eviction from the LLC. Such addresses are effectively already *dead* in the LLC at their time of eviction. Therefore, their eviction by inaccurate prefetches *does not cause pollution.* This class comprises a dominant 84% of all LLC victim addresses even for a small 2MB LLC.
- **PrefetchedBeforeUse**: LLC victim addresses that are prefetched into on-die caches before their next access by a demand. The eviction of such LLC victim addresses by inaccurate prefetches does increase memory traffic but *does not* not cause cache pollution. This class comprises nearly 13% of all LLC victim addresses for a 2MB LLC.
- **BadPollution**: LLC victim addresses whose next demand access misses in on-die caches and goes to main memory. These addresses are the true victims of pollution by inaccurate prefetches. However, this class comprises only 3% of LLC victim addresses even for a small 2MB LLC.

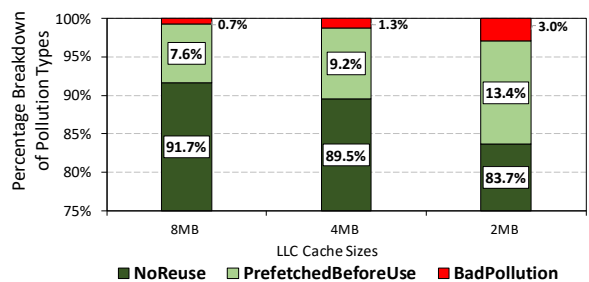

Figure 20: Breakdown of the types of cache pollution caused by inaccurate prefetches. The y-axis starts at 75%.

We conclude that the pollution impact of inaccurate prefetches is relatively small in the configurations and the workloads we examine.

Multiple prior works [39, 44, 46, 53, 70, 74, 79] also report similar observations on the abundance of dead cache lines in LLC in the presence of an aggressive prefetcher. These works either determine the fill/replacement priority of a prefetched cache line based on prefetcher accuracy or design dead-block predictors to identify dead lines as high-priority replacement candidates.